\documentclass[twocolumn,showpacs,preprintnumbers,amsmath,amssymb,floatfix]{revtex4}

\usepackage{graphicx}

\usepackage{dcolumn}

\usepackage{bm}

\def\stacksymbols #1#2#3#4{\def\theguybelow{#2}
        \def\verticalposition{\lower#3pt}
        \def\spacingwithinsymbol{\baselineskip0pt\lineskip#4pt}
        \mathrel{\mathpalette\intermediary#1}}

\def\intermediary#1#2{\verticalposition\vbox{\spacingwithinsymbol
        \everycr={}\tabskip0pt
        \halign{$\mathsurround0pt#1\hfil##\hfil$\crcr#2\crcr
                \theguybelow\crcr}}}

\begin{document}

\preprint{PRL}

\title{Self-Similar Evolution of Parabolic Pulses in a Laser }

\author{F. \"{O}. Ilday$^{1,*}$, J. R. Buckley$^{1}$, W. G. Clark$^{2}$, and F. W. Wise$^{1}$}

\affiliation{$^{1}$Department of Applied Physics, Cornell University, 212 Clark Hall, Ithaca, NY 14853, USA.\\ 
$^{2}$Clark-MXR Inc., Dexter, MI 48130.\\
$^{*}$Corresponding author. Electronic address: ilday@ccmr.cornell.edu}

\begin{abstract}

Self-similar propagation of ultrashort, parabolic pulses in a laser resonator is observed theoretically and experimentally. This constitutes a new type of pulse-shaping in modelocked lasers: in contrast to the well-known static (soliton-like) and breathing (dispersion-managed soliton) pulse evolutions, asymptotic solutions to the nonlinear wave equation that governs pulse propagation in most of the laser cavity are observed. Stable self-similar pulses exist with energies much greater than can be tolerated in soliton-like pulse shaping, and this will have implications for practical lasers.

\end{abstract}


\maketitle

\noindent

Self-similarity is a recurring theme in the description of many natural phenomena. The emergence of self-similarity in complicated, nonlinear systems can be particularly informative about the internal dynamics: self-similarity arises after the influences of initial conditions have faded away, but the system is still far from its ultimate state~\cite{SS-int-asym}.  Furthermore, the presence of self-similarity implies an inherent spatial and/or temporal order that can be exploited in the mathematical treatment of the governing equations.  For example, reduction-of-symmetry techniques effectively reduce the dimensionality of the system,~\cite{symmetry-reduction} which facilitates analysis. In the field of nonlinear optics, a limited number of self-similar (SS) phenomena have been reported. As examples, SS behavior in stimulated Raman scattering ~\cite{SS-SRS} and the formation of Cantor-set fractals in materials that support spatial solitons ~\cite{Cantor} were investigated.  The most pertinent prior work is the theoretical demonstration of SS propagation of short pulses of parabolic intensity profile in optical fibers with normal group-velocity dispersion (GVD) and strong nonlinearity~\cite{wave-breaking-free}. Recently, this concept has been extended to an optical fiber amplifier.  By use of the technique of symmetry reduction, parabolic pulses were shown to propagate self-similarly in the amplifier, and the predicted evolution was verified experimentally~\cite{parabolic-amp1,parabolic-amp2,parabolic-amp3}. Considering that the solutions of the nonlinear Schrodinger equation that governs pulse propagation in a fiber are well-known, it is remarkable that the SS solutions were discovered only recently.

In addition to the scientific interest in SS pulse propagation, there is practical motivation: the energy of ultrashort optical pulses is generally limited by wave-breaking, which is a consequence of excessive nonlinearity. The limitation is particularly stringent in short-pulse fiber devices, where small modes produce high intensities and therefore large nonlinear phase shifts.  In contrast to solitons, SS pulses (or ``similaritons''~\cite{Raman-similariton}) can tolerate strong nonlinearity without wave-breaking.  The presence of normal GVD tends to ``linearize'' the phase accumulated by the pulse, which increases the spectral bandwidth but does not destabilize the pulse~\cite{wave-breaking-free}. Owing to their compact size and excellent stability, fiber lasers are appealing alternatives to bulk solid-state lasers. If the pulse energy available from fiber lasers can be increased to that of solid-state lasers, fiber lasers could find widespread application, so the possibility of achieving SS propagation in a laser is very attactive.

However, experimental studies of SS optical phenomena are generally restricted to the observation of evolution over a limited number of characteristic propagation lengths, due to practical limitations. Furthermore, we are not aware of any experimental observations of asymptotic SS evolution in an optical system with feedback, such as a laser. Asymptotic solutions of a wave equation would appear to be incompatible with the periodic boundary condition of a laser resonator.

Here we report numerical predictions of the existence of stable pulses that indeed propagate self-similarly in a laser cavity, along with clear experimental evidence of SS pulses in a fiber laser. This regime of operation constitutes a new type of pulse shaping in a modelocked laser, qualitatively distinct from the well-known soliton~\cite{soliton-fiber-laser} and dispersion-managed soliton~\cite{DM-soliton} regimes. The main features of SS pulses in a laser will be presented, and the implications for practical devices will be discussed briefly.

Pulse formation in a femtosecond laser is typically dominated by the interplay between dispersion and nonlinearity ~\cite{SS_Haus_theory}. An effective saturable absorber (SA) is required for initiation of pulsed operation from intra-cavity noise and subsequent stabilization of the pulse. Hence, the laser constitutes a dissipative system and its basic features can be understood within the formalism of a complex Ginzburg-Landau equation or related formulations~\cite{CGLE}.

Soliton fiber lasers are limited to low pulse energies (100 pJ or less)~\cite{soliton-fiber-laser}. At higher energies, nonlinear effects cause wave-breaking~\cite{wave-breaking}, which leads to multiple-pulsing (more than one pulse circulates in the cavity). The pulse can tolerate only a small nonlinear phase shift ($\Phi^{NL} << \pi$) before such instabilities occur. Stretched-pulse fiber lasers consist of segments of anomalous and normal GVD~\cite{DM-soliton}. These implement the concept of dispersion management (DM)~\cite{DM},  and support the analog of DM solitons. DM solitons can tolerate nonlinear phase shifts an order of magnitude larger than ordinary solitons, and the pulse energy thus exceeds the soliton energy by the same factor.

Solitons are static solutions of a nonlinear wave equation, and DM solitons are breathing solutions.  In contrast, SS pulses are asymptotic solutions to the governing equation. The evolution of their properties (e.g., the pulse duration) is monotonic. Such a solution cannot be a stable solution in a cavity; a mechanism to reverse any changes and thus restore the solution after traversal of the cavity must be provided.  Another issue must be addressed for SS propagation to occur in a laser: in general, pulses will evolve to fill available gain bandwidth. However, self-similar propagation of intense pulses is disrupted if the pulse encounters a limitation to its spectral bandwidth~\cite{Dudley_bw_similariton}.

\begin{figure}

\centerline{\includegraphics[width=8.5cm]{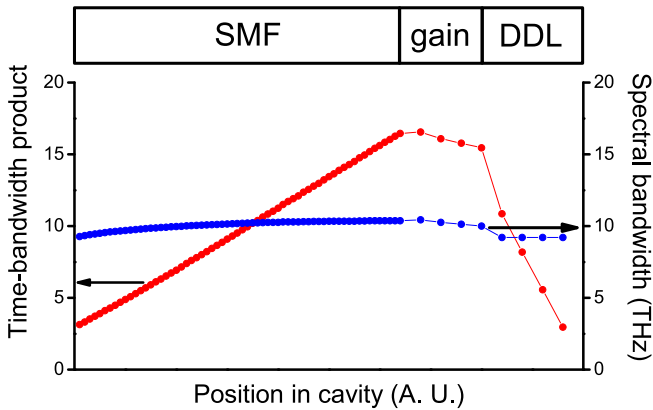}}

\caption{\label{similariton_setup} Top: schematic diagram of the main components of the laser. SMF: single mode fiber, DDL: dispersive delay line, SA: saturable absorber. The cavity is a ring, so after the DDL the pulse goes into the SMF. Bottom: variation of the time-bandwidth product and spectral width in the cavity. }

\end{figure}

The conceptual model for a fiber laser illustrated in Fig.~\ref{similariton_setup} addresses these issues. A segment of single-mode fiber (SMF) with normal GVD forms the majority of the cavity. Amplification is provided by the minimum length of gain fiber required to provide adequate gain. The pulse will experience negligible GVD and nonlinearity during amplification, which effectively decouples bandwidth filtering from the nonlinear evolution in the SMF. The gain fiber is followed by a SA, which also serves as the output coupler. The final element is a dispersive delay line (DDL) that provides anomalous GVD with negligible nonlinearity. The cavity is a ring: after the DDL, the pulse returns to the SMF. In a time-domain view, chirp accumulated in the desired SS propagation will be compensated by the DDL.  In the frequency domain, generated bandwidth will be filtered by the gain medium, to produce a self-consistent solution in the resonator. The pulse evolution is illustrated by the plots of time-bandwidth product and spectral width versus cavity position (Fig.~\ref{similariton_setup}, details to be provided below).

Numerical simulations of the model laser were employed to search for SS solutions, and to verify that such solutions are accessible from intracavity noise.  The parameters of the numerical model correspond to the experimental setup. A segment of SMF $6$ m in length is connected to a 23-cm-long gain fiber, and a grating pair is used for the DDL. Propagation within each section is modeled by 
\begin{eqnarray}
\frac{\partial A(\xi,\tau)}{\partial \xi} + i \beta_{2}\frac{\partial^2 }{\partial \tau^2} A(\xi,\tau) = \nonumber\\i \gamma |A(\xi,\tau)|^{2} A(\xi,\tau) + g(E_{pulse}) A(\xi,\tau),
\end{eqnarray}
where $A(\xi,\tau)$ is the envelope of the field, $\xi$ is the propagation coordinate, $\tau$ is time scaled to the pulse duration, $\beta_{2} = 23 fs^2/mm$ is the GVD, and $\gamma = 0.0047 (Wm)^{-1}$ is the coefficient of cubic nonlinearity for the fiber section. The pulse energy is $E_{pulse} = \int^{T_{R}/2}_{-T_{R}/2}{|A(\xi,\tau)|^2 d\tau}$ and $T_{R} \sim 40$ ns is the cavity roundtrip time. $g(E_{pulse})$ is the net gain, which is non-zero only for the amplifier fiber. The gain saturates with total energy according to
\begin{equation}
g(E_{pulse}) = \frac{g_{0}}{1 + E_{pulse}/E_{sat} }, 
\end{equation}
where $g_{0} \sim 30$ dB is the small-signal gain with an implicit parabolic frequency dependence and a bandwidth of 45 nm. The gain saturation energy, $E_{sat}$, is determined by the pump power. The SA is modeled by a transfer function that describes its (monotonic) transmittance $T = 1 - l_{0}/(1 + P(\tau)/P_{sat})$, where $l_{0} = 0.3$ is the unsaturated loss, $P(\tau)$ is the instantaneous pulse power and $P_{sat}$ is the saturation power. All significant linear losses occur in the DDL and in coupling into the SMF. The pulse amplitude is reduced by a factor of $\sim 10$ after the SA to account for these losses, which can be lumped together without loss of generality.

\begin{figure}

\centerline{\includegraphics[width=8.5cm]{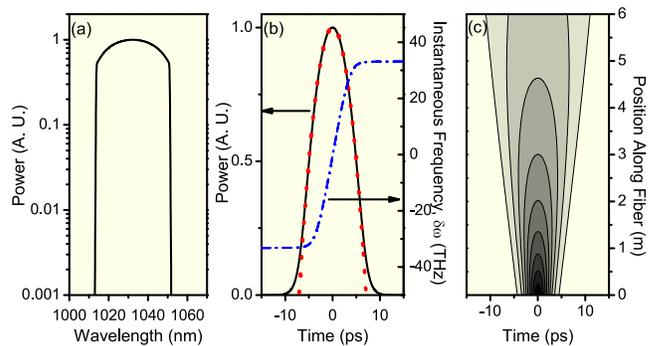}}

\caption{\label{numerical}  Results of numerical simulations. (a) Output power spectrum. (b) Temporal intensity profile (solid curve), instantaneous frequency (dash-dotted curve), and a parabolic fit (dotted curve) to the intensity profile. (c) Equal-intensity contours of the pulse evolution within the SMF.}

\end{figure}

\begin{figure}

\centerline{\includegraphics[width=7.5cm]{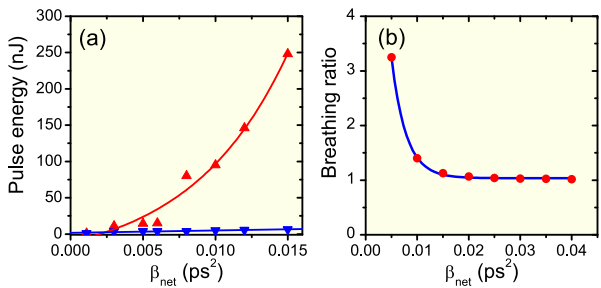}}

\caption{\label{similariton_energy}  (a) Calculated maximum pulse energy {\em versus} total dispersion for the self-similar (up-triangles) and stretched-pulse (down-triangles) schemes. (b) Intra-cavity breathing ratio with respect to total dispersion for the self-similar scheme. The lines are to guide the eye. }

\end{figure}

The numerical model is solved with a standard symmetric split-step beam propagation algorithm and the initial field is white noise. The simulations are run until the field becomes constant after a finite number of traversals of the cavity. We have verified numerical accuracy by checking that the results are unchanged after doubling each of the sampling resolutions. The simulations exhibit stable pulse formation over a reasonably-wide range of parameters. For fixed parameters, exactly the same stable solutions are reached from distinct initial noise fields. Thus, these solutions appear to be true attractors. A typical solution obtained with ${\rm \beta_{net}} = 0.007\;{\rm ps^2}$, $P_{sat} = 15$ kW, and $E_{sat} = 4.0$ nJ is shown in Fig.~\ref{numerical}. The pulse energy is 15 nJ. The approximately-parabolic temporal profile, nearly-linear chirp, shape of the power spectrum, and the evolution of the intensity profile are all signatures of SS pulse formation. For increasing normal GVD (${\rm \beta_{net}} > 0$) the pulse energy increases dramatically (see Fig.~\ref{similariton_energy}). The stretching ratio decreases exponentially at fixed pulse energy as the net GVD is increased (inset of Fig.~\ref{similariton_energy}). Thus, the energy-scaling can be understood intuitively:  the (exponential) increase in pulse energy with increasing GVD maintains a fixed stretching ratio.

Encouraged by the simulation results,  we constructed a Yb:fiber laser with the parameters listed above. Mode-locked operation is initiated and stabilized by nonlinear polarization evolution (NPE)~\cite{similariton_NPE}, and the output is taken from the NPE rejection port. The laser mode-locks easily upon adjustment of the wave-plates that control the NPE. The experiments are performed at low pulse energies ($\sim 2$ nJ) to avoid over-driving NPE and thus isolate the SS pulse-shaping. Results obtained with ${\rm \beta_{net}} = 0.004\;{\rm ps^2}$ will be presented and are representative of those obtained in the range ${\rm \beta_{net}} = 0.003 - 0.015\;{\rm ps^2}$.

\begin{figure}

\centerline{\includegraphics[width=8.5cm]{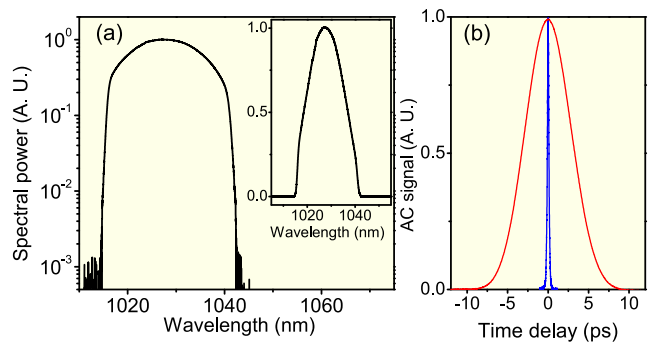}}

\caption{\label{experimental} Experimental results. (a) Power spectrum on semi-log scale. Inset: spectrum on linear scale. (b) Intensity autocorrelation of the chirped and the dechirped pulses. } 

\end{figure}

The characteristics of the generated pulses are consistent with SS evolution. The power spectrum exhibits the unique shape that is predicted numerically: parabolic near the peak, with a transition to a steep decay (Fig.~\ref{experimental} (a)).  The pulse is extracted from the cavity where it is expected to have maximum positive chirp. The intensity autocorrelation reveals that the pulse duration is $\sim 4.2$ ps. After dechirping the pulse with diffraction gratings external to the cavity, the pulse duration is 130 fs (Fig.~\ref{experimental} (b)). The dechirped pulse duration is close to the transform-limit, which implies an approximately linear chirp on the generated pulse, as expected theoretically.

The magnitude of anomalous GVD required to dechirp the output pulse is an important diagnostic of pulse evolution in the laser.  From the experimental value ($- 0.170\;{\rm ps^2}$, approximately equal to that provided by the intracavity gratings) we conclude that the pulse chirp is positive throughout the cavity, with a minimum at the beginning of the SMF. We estimate a minimum pulse duration of 250 fs. Thus, the overall stretching ratio {\em within} the cavity is $\sim 17$.

\begin{figure}

\centerline{\includegraphics[width=8.5cm]{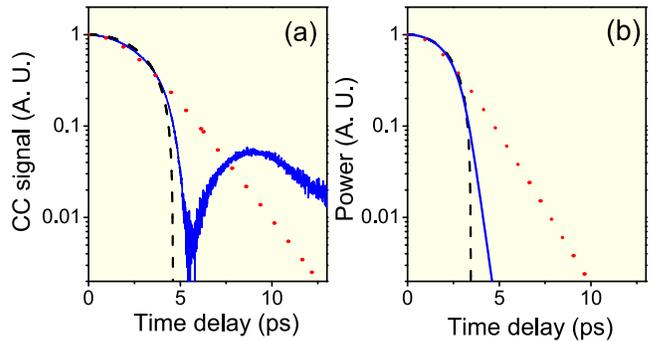}}

\caption{\label{CC} (a) Cross-correlation of the chirped and the dechirped pulses. (b) Calculated intensity profile. Parabolic (dashed curves) and ${\rm sech^2}$ (dotted curves) fits are also shown.}

\end{figure}

Finally, the measured shape of the chirped pulses generated by the laser further supports the SS nature of the pulse evolution. The pulse shape was obtained by cross-correlation with the dechirped pulse, which is $\sim 32$ times narrower and thus a good approximation to a delta-function. The pulse shape (Fig.~\ref{CC}) agrees well with the numerical simulations and the parabolic intensity profile expected analytically. The small secondary pulse could develop during intermediate asymptotic evolution~\cite{parabolic-amp3}. However, it does not appear in the simulations. Qualitatively, such structure is expected when the pulse is taken from the NPE rejection port. The simulated results shown above are for the pulse that continues to propagate in the cavity, so this minor discrepancy is not surprising.

The pulse energy was intentionally limited as a control on the experiments reported above.  An important aspect of the SS regime is the prediction of stable pulses with much higher energies than can be contained in DM solitons. This prediction is based on the assumed monotonically-saturating absorber. NPE is interferometric in nature, so it can be overdriven (i.e., the transmission eventually decreases at high intensity) and this may well limit the pulse energy. A laser with lower repetition rate was constructed, so that high-energy pulses were possible energetically given the available pump power.  With ${\rm \beta_{net}} = 0.003\;{\rm ps^2}$, stable pulses with autocorrelation and spectrum similar to those shown above are generated.  The chirped-pulse energy is 10 nJ, which is 3 times the maximum obtained from a DM-soliton laser~\cite{3nJ-SPL}. More systematic studies of the SS pulse energy are needed, but this initial result is encouraging. A laser operated close to the SS regime has exhibited wave-breaking-free operation with peak powers 5 times that of a DM-soliton laser~\cite{OmerWBF}.

The  experimental results confirm the essential features of the SS regime of laser operation, which can be placed in context with prior lasers.  In contrast to static  soliton solutions and breathing DM solitons, the nonlinear pulse evolution is monotonic in each segment of the dispersion map.  DM solitons stretch and compress twice per cavity round-trip, while SS pulses stretch and compress once.  True solitons are unchirped. DM solitons acquire both negative and positive chirps as they propagate, while SS pulses are positively-chirped throughout the cavity. Detailed distinctions include the approximately parabolic temporal profile, which contrasts with the hyperbolic-secant soliton solution and gaussian DM soliton, and the unique spectral profile of the SS pulse.

In this report we have focused on fiber lasers, which perhaps provide the most natural environment for SS pulses.  However, the concept is more broadly applicable. SS operation of solid-state lasers would extend the previously-observed operation with normal GVD ~\cite{posGVDTi_sapph}, where the pulse duration is nearly static, to stretching ratios $> 1$. Initial simulations with parameters close to those of existing Ti:sapphire lasers that produce $\sim$10-fs pulses exhibit some of the features of SS propagation (including stable chirped solutions and increased pulse energy) when the net GVD is normal. It is also reasonable to expect that the conditions for SS pulse formation could be met in semiconductor lasers.

In conclusion, we have observed self-similar pulse evolution in a laser resonator. This new type of pulse formation is qualitatively distinct from previously-identified mode-locking techniques. Chirped pulses with parabolic temporal (and spectral) profiles exist throughout the laser cavity. The pulses can be dechirped to nearly the Fourier-transform limit outside of the laser, so the demonstration of self-similar pulse formation may have major implications for short-pulse lasers. The initial experimental results reported here already represent at least twice the pulse energy of prior femtosecond fiber lasers. If overdriving of the effective saturable absorber can be avoided, pulse energies one to two orders of magnitude larger than those of existing lasers should be possible. Current efforts are aimed at identifying the limits of NPE and exploring monotonic saturable absorbers.

This work was supported by the National Institutes of Health under grant EB002019 and the Air Force Research Laboratory under contract F29601-00-C-0049.

\end{document}